\input epsf.tex

\magnification=\magstep1

\hsize=17 truecm
\vsize=25 truecm
\hoffset = -0.34 truecm
\voffset = -1.04 truecm
\centerline{\bf A Precision Test Of Decoherence}

\vskip 2\baselineskip

\centerline{J.R. Anglin and W.H. Zurek}

\centerline{Theoretical~Astrophysics, T-6, Mail~Stop~B288}
 
\centerline{Los~Alamos~National~Laboratory}

\centerline{Los~Alamos, New~Mexico~87545}
\vskip\baselineskip

\noindent ABSTRACT: The motion of a charged particle over a conducting
plate is damped by Ohmic resistance to image currents.  This
interaction between the particle and the plate must also produce
decoherence, which can be detected by examining interference patterns
made by diffracted particle beams which have passed over the plate.
Because the current densities within the plate decay rapidly with the
height of the particle beam above it, the strength of decoherence
should be adjustable across a wide range, allowing one to probe the
full range of quantum through classical behaviour.

\vskip 2\baselineskip\centerline{\bf Introduction}

Charged particles passing close to conducting surfaces have been
investigated in tests of the Ahoronov-Bohm effect $^1$) and the
equivalence principle $^2$).  The phenomenon of dissipation
due to electrical resistance to image currents in the conductors has
been investigated extensively, but the effects on the particles of the
concomitant resistor noise have been discounted $^3$).  Among
these effects will be decoherence$^{4}$).  We propose that the strong
dependence of the image current densities to the easily varied
parameter of trajectory height makes the motion of a charged particle
over a conducting plate an excellent precision test for our
understanding of decoherence.

\epsfxsize = 16 truecm
\vbox{\vskip 3 truecm \epsfbox[70 250 530 480]{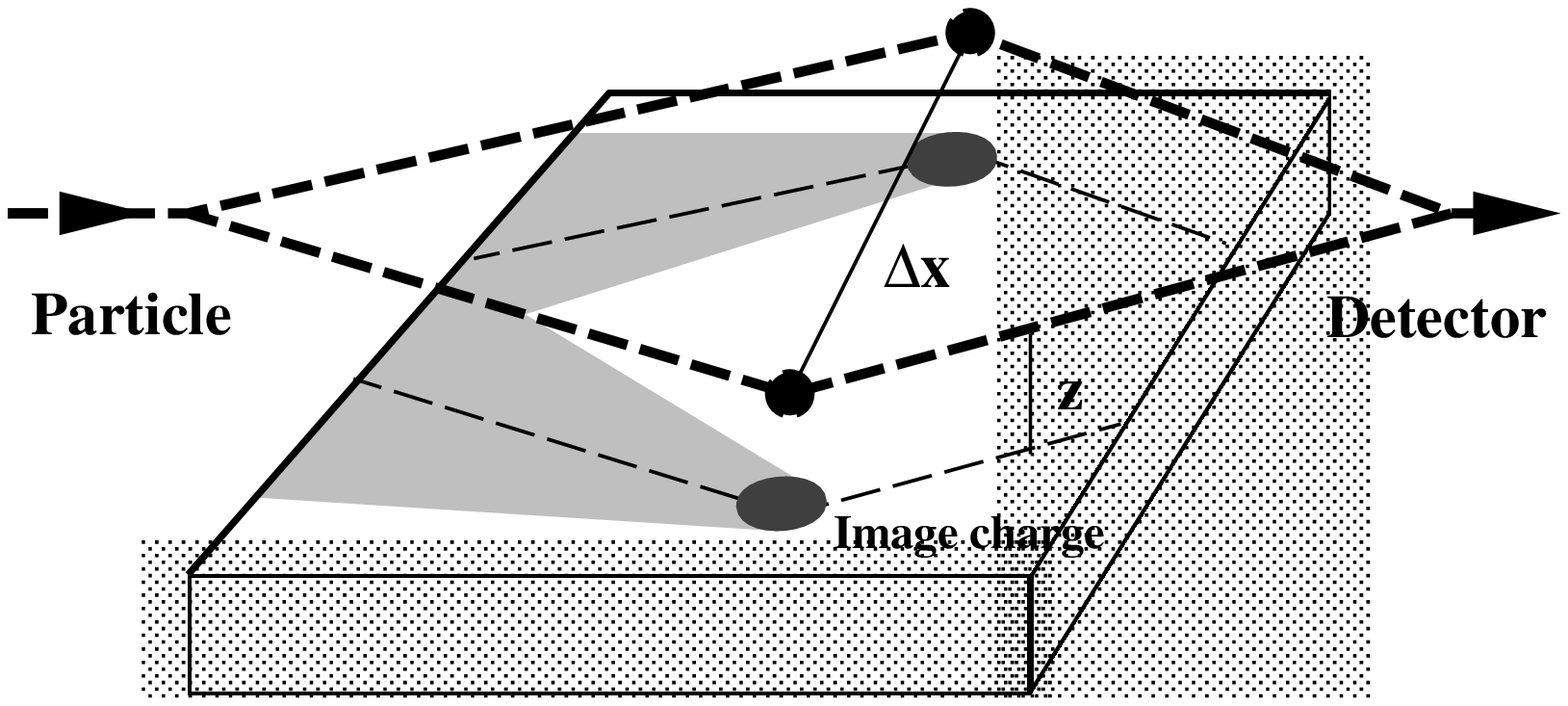}}

\vskip -1 truecm


\parshape= 1 1 truecm 15 truecm 
\noindent{\bf Figure 1:~}
Sketch of proposed system.  The heavy dashed lines indicate two
trajectories of the particle over the conducting plate.  The large shaded
regions represent the disturbance in the electron gas inside the plate.


\centerline{\bf Decoherence}

If a charged particle passes over a conducting plate along some
trajectory, it induces an image charge on the surface of the conductor,
which moves with the particle by inducing bulk currents within the body
of the conductor $^5$).  These bulk currents encounter Ohmic resistance,
which is primarily due in metals at room temperature to the scattering
of coherent electron flow by thermal phonons.  The passage of the
charged particle along a trajectory over the conductor thus leaves a
wake of disturbance in the electron and phonon gas beneath it.

After the charge has crossed the plate (and continued on its flight
towards a detector), the disturbed state of the conductor remains 
as a record of the charge's passage (and in fact, of the
particular path that it took).  Quantum mechanically, this means that
the state of the electron gas and of the charged particle must be
described together by a joint wave function.  The contribution to this
wave function from a single trajectory $\cal{T}$, at the end of which
the particle is at position $Q$, can be written as
$$
|\Psi_{\cal T}\rangle =  |Q\rangle \,
|\psi_{E}[{\cal T}]\rangle\;.\eqno
$$

According to Feynman's picture of quantum evolution, the final state of
the system is found by a weighted sum of such states $|\Psi_{\cal
T}\rangle$ over all trajectories ${\cal T}$ which the particle may have
taken.  Quantum mechanics then predicts the probability of finally
observing the particle at a given position $Q$ (without measuring any
properties of the conducting plate) to be
$$
\eqalign{
P(Q) &= \sum_{Q_i,Q_i'}\phi(Q_i)\phi^*(Q_i')\cr
&\qquad\times \sum_{\cal T}\sum_{\cal T'} 
	e^{{i\over\hbar}(S[{\cal T}] - S[{\cal T'}])}
	\langle \psi_{E}[{\cal T'}]|\psi_{E}[{\cal T}]\rangle\;,}
\eqno
$$
where the sums are over trajectories ${\cal T}$ or ${\cal T'}$ in which
the particle begins at $Q_i$ or $Q_i'$ and ends up at $Q$, and
$\phi(Q_i)$ is the wave function describing the initial state of the
particle.

If this initial state is a superposition of two distinct localized
states, such as is produced by a double slit, cross terms in (99)
can produce a probability distribution $P(Q)$ which is not the
classical ``two lumps'', but an interference pattern characteristic of
wave mechanics.  The familiar textbook case in which this possibility
is fully realized occurs only when $\langle \psi_{E}[{\cal
T'}]|\psi_{E}[{\cal T}]\rangle \to 1$, so that the conducting plate
does not effectively distinguish between different particle
trajectories.  In general, though, this inner product is {\it not}
unity, and trajectories that are far apart may disturb
the plate's electron gas in such different places that the inner
product of the electron states will be negligible.  Interference
between these trajectories will therefore be suppressed, and the final
probability distribution of the particle will thus be altered towards
the classical limit.  

This phenomenon is an example of the process known as {\it
decoherence}$^4$), which is believed to play a crucial role in enforcing the
effective classicality of macroscopic physics, and is thought to be the
greatest challenge facing such hypothetical advanced technologies as
quantum computing$^6$).  A precision test of decoherence is thus highly
desirable.

\vskip\baselineskip\centerline{\bf Precision Test of Decoherence}

The system of charged particle and conducting plate is suitable for
a precision test of decoherence, because the strength of their
interaction is adjustable over a wide range.  Calculations using
classical electromagnetism, for a charge $Q$ moving over the plate at a
height $z$ and constant velocity $v$ show a rate of Joule heating in
the conductor which is  
$$
P_C = {Q^2\rho v^2\over 16\pi z^3}\;,\eqno
$$
where $\rho$ is the specific resistivity of the plate.  This implies an
Ohmic dissipation rate proportional to $d^{-3}$: we can compute the
relaxation time $\tau_r$ to be
$$
\tau_r = {v\over\dot{v}}\sim \Bigl({z\over 10^{-4} {\rm m}}\Bigr)^3 \times 
\cases{
2\times 10^{3}{\rm s}\; & electron\cr
		3{M\over Q^2}\times 10^{6}{\rm s}\; & ion\cr}
\eqno
$$
where $M$ is the ion mass in units of proton mass, and we take the
resistivity $\rho$ to be on the order of $10^{-6} \Omega {\rm m}$.  (This
rather high resistivity is possessed by manganese at room temperature;
since a resistivity a few orders of magnitude higher still could
be even more convenient, alloys or semi-conductors might be contemplated
instead of a pure metal.)

(Note that adding a thin layer of insulator on top of the conducting
plate will actually produce dissipation proportional to the inverse
{\it fourth} power of $z$ $^{2,5}$).  The sensitivity to $z$ that
we are considering can thus be made even greater; but we will not
discuss this option here in detail.)

These calculations are classical, and a proper quantum treatment is
certainly required to make serious theoretical predictions concerning
decoherence in this system.  We must consider the unitary evolution of
phonons and a nearly free electron gas as a charge passes over the
plate on an arbitrary trajectory.  The electromagnetic field may
presumably be treated classically, and for sufficiently slow particles
an adiabatic expansion to first order in the particle speed will
probably suffice.  After the charge has crossed the plate, the direct
perturbation of the conductor by the charge will cease, but the
resistive interactions among the conductor's constituents that were
driven by the charge, as it dragged an image current of electrons
through the mill of thermal nuclei, will have
brought the conductor into a final quantum state that depends on the
particle's trajectory.  The inner product between two such states must
then be computed.

For the present brief communication, we merely suppose that the full
quantum calculation will recover both the classical dissipation rate,
and the result (typical in linear systems) that decoherence rates are
proportional to it.  As a zeroth order approximation that should be
reasonable at room temperature, we will assume that the decoherence
time scale $\tau_d$ is given by the formula for a completely linear
model at high temperature$^7$),
$$
\tau_d = \tau_r \Bigl({\lambda_{dB}\over\Delta x}\Bigr)^2\;,\eqno
$$
where $\lambda_{dB} \equiv h/\sqrt{2mk_B T}$ is the thermal de Broglie
wavelength, and $\Delta x$ is a length scale characterizing the
difference between the two quantum states that are to decohere.  In our
case, we will take this scale to be the distance between Feynman
trajectories across the plate, which could perhaps be of order $10^{-4}
{\rm m}$.  

Since $\tau_r$ is proportional to the particle mass, we therefore
estimate the same decoherence time for singly charged ions and
electrons:
$$
\tau_d \sim \Bigl({z\over 10^{-4} {\rm m}}\Bigr)^3 \times 10^{-5} 
{\rm s}\;.\eqno
$$
If we assume that our charged particles will traverse a conducting
plate that measures a centimeter across, flying at a speed on the order
of a kilometers per second, the decoherence time will be on the order
of the time of flight for $z\sim 0.1$ mm.  By varying the trajectory
height $z$ we should therefore indeed be able to probe the full range
from negligible to strong decoherence.

\centerline{\bf Conclusion}

There are a number of aspects of decoherence which one would like to
have clarified by a precision experiment.  One of the most interesting
is the dependence of the decoherence of two trajectories on the spatial
distance between them.  In simple linear models, decoherence has a
Gaussian profile with distance, but in general we expect that this
profile will instead approach some non-zero constant value at large
distances $^{8,9}$).  In the particle and conductor
system, we expect the length scale at which this saturation occurs to be
set by the correlation length in the electron gas.

By varying the separation between the two slits through which the
particle is coherently passed before crossing the plate (or using
different crystal lattices for diffraction, if a low intensity electron
beam is used instead of a stream of ions), one can vary the distances
over which quantum interference of trajectories is important in
determining the final $P(Q)$.  This should allow one to probe the
dependence of decoherence on spatial separation.  By also varying the
overall strength of decoherence, through the trajectory height $d$, we
can investigate decoherence in a two dimensional parameter space, and
provide a stringent test of decoherence in a system which is
theoretically non-trivial.  Unlike set-ups deliberately designed to
mimic simple theoretical models for decoherence, the particle crossing
a conducting plate will thus be able to stretch our theory: it is a
real precision test of decoherence.

\vskip\baselineskip\centerline{\bf References}


\noindent [1] T.H. Boyer, Phys. Rev. D {\bf 8}, 1667 (1973); {\bf 8}, 1679 (1973).

\noindent [2] T.W. Darling {\it et al.}, Rev. Mod. Phys. {\bf 64}, 237 (1992).

\noindent [3] L.H. Ford, Phys. Rev. D {\bf 47}, 5571 (1993).

\noindent [4] W.H. Zurek, Phys. Rev. D {\bf 24}, 1516 (1981); {\it Physics Today}, October 1991.

\noindent [5] T.H. Boyer, Phys. Rev. A {\bf 9}, 68 (1974).

\noindent [6] I.L. Chuang, R. Laflamme, P. Shor, and W.H.~Zurek,
{\it Science} {\bf 270}, 1633 (1995).

\noindent [7] W.H. Zurek, in G.T. Moore and M.O.~Scully, eds., {\it
Frontiers of Nonequilibrium Statistical Physics} (Plenum: New York,
1986).

\noindent [8] M.R. Gallis and G.N. Fleming, Phys. Rev. A {\bf 42}, 38 (1990); {\bf 43}, 5778 (1991).

\noindent [9] J.R. Anglin, J.P. Paz, and W.H. Zurek, ``Deconstructing Decoherence'', in preparation.

\vfill

\end